 \documentclass[preprint]{aastex}

\slugcomment{\hskip -25 pt To appear in Ap. J. Letters. Subm. 2001 Mar 10; rev.
2001 Apr 16.}

\shortauthors{Fern\'andez, Jewitt \& Sheppard}
\shorttitle{Low Albedos Among Extinct Comet Candidates}

\begin{document}

\title{Low Albedos Among Extinct Comet Candidates}

\author{Yanga R. Fern\'andez\altaffilmark{1}, David C. Jewitt\altaffilmark{1},
Scott S. Sheppard}
\affil{Institute for Astronomy, Univ. of Hawai`i at M\=anoa, \\
2680 Woodlawn Dr., Honolulu, HI 96822}
\email{yan@ifa.hawaii.edu, jewitt@ifa.hawaii.edu, sheppard@ifa.hawaii.edu}

\altaffiltext{1}{Visiting Astronomer at W. M. Keck Observatory, which is 
jointly operated by the California Institute of Technology and the 
University of California.}

\begin{abstract}

We present radiometric effective radii and visual geometric albedos for
six asteroids in comet-like orbits. Our sample has three of the four
known retrograde asteroids (1999 LE$_{31}$, 2000 DG$_8$, and 2000
HE$_{46}$) and three objects ((18916) 2000 OG$_{44}$, 2000 PG$_3$, and
2000 SB$_1$) on prograde but highly elliptical orbits. These measurements
more than double the number of known albedos for asteroids with a
Tisserand invariant in the cometary regime. We find that all six of our
objects, and nine of the ten now known, have albedos that are as low as
those of active cometary nuclei, which is consistent with their
supposed evolutionary connection to that group. This albedo
distribution is distinct from that of the whole near-Earth
and unusual asteroid population, and the strong correlation between
Tisserand invariant and albedo suggests there is a significant
cometary contribution to this asteroid population.

\end{abstract}


%%%%%%%%%%%%%%%%%%%%%%%%%%
\section{Introduction}

An old comet that loses all of its available volatiles or is covered by
a mantle that prevents sublimation of subsurface ice will
observationally appear as a near-Earth or unusual \footnote{The IAU
Minor Planet Center currently defines ``near-Earth'' asteroids as those
with perihelion distances below 1.3 AU, and ``unusual'' asteroids as
those that are not near-Earth and residing neither in the Main Belt nor
wholly in transjovian space.} asteroid (NEA or UA). The dynamical
lifetime of short-period comets are about 10 to 100 times longer than
the devolatilization timescale \citep{ld94}, so one expects to see such
extinct comets if they do not disintegrate or collide with a planet.
However there is currently no way to determine if a given asteroid is such
a comet (short of witnessing last gasp activity), so we do not
yet know the fraction of extinct comets in the asteroid population.
Knowing this fraction would give clues to the physical evolution of
comets and the hazard to Earth from asteroid and comet collisions.

Dynamical models of the cometary component are rendered inconclusive by
the unknown extent and duration of nongravitational forces on active nuclei
\citep{hb98}. Some models \citep{bot00} can even fit
the NEA population without a cometary source.
Nevertheless, the existence of the
now-inactive comet 107P/Wilson-Harrington, which has been observed to
have a coma just once in the 51 years since discovery
\citep{bow92,fer97}, strongly suggests that there is a non-zero
cometary contribution, and we have taken an observational tack to
address the problem.  A statistical indicator for the cometary origin
of an asteroid comes from the albedo and orbit shape. All known albedos
of cometary nuclei are low and almost all inner Solar System cometary
orbits have common characteristics. Thus an asteroid in a comet-like
orbit with a comet-like albedo is a good candidate extinct comet.

Nearly a dozen albedos of active cometary nuclei have been measured and
all are dark, with geometric albedos ranging from 0.02 to 0.12
\citep{jew92,fer99}. Thus we expect to see similar albedos among the
extinct comet fraction of NEAs and UAs. Albedo alone is not a unique
determinant of origin however since there are many non-cometary,
dark asteroids in the outer Main Belt \citep{gra89} from which some
NEAs and UAs could have come.

The Tisserand invariant $T_J$ \citep{tis96}, a constant of motion in
the restricted three-body problem, can be used to separate objects by
dynamical class. The threshold $T_J=3$ separates objects 
coupled to ($<3$) or decoupled from ($>3$) Jupiter.  Generally
asteroids have $T_J>3$, Jupiter-family (JF) comets have $2<T_J\le3$,
and Halley-family (HF) and long-period (LP) comets have $T_J\le2$
\citep{lev96}, though the scheme is not fail-safe.  There are currently
(mid-April 2001) 11 asteroids with $T_J<2$, 4 of which are in
retrograde orbits, and 131 NEAs and UAs with $2<T_J<3$. The total
number of NEAs and UAs is currently 1,400, so 10\% have $T_J<3$. Among
NEAs alone, 82 out of 1,327, or 6\%, have $T_J<3$.  In this Letter we
describe the radiometric determination of the albedos and effective
radii of six $T_J<3$ asteroids. The Tisserand values, the contributing
orbital elements, and the geometry of our observations are given in
Table 1. The new data bring to ten the total number of $T_J<3$
asteroids with known albedos.

%%%%%%%%%%%%%%%%%%%%%%%%%%
\section{Observations and Reduction}

The observations span two wavelength regimes, mid-infrared
(MIR) and visible (simultaneously for 4 of the 6 objects). The 
MIR images were obtained with the Keck I
telescope using the LWS array \citep{jp93} in June 2000 and with the
Keck II telescope using the MIRLIN array \citep{res94} in November 2000. The
visible data were obtained with the Univ. of Hawaii 2.2-m
telescope using a Tek2048 CCD in July and November 2000.  Table 2 gives
the averages of the measured flux densities. All objects were point sources.

The MIR data were obtained using chopping and nodding, with throws of 4
arcsec. Non-sidereal guiding was used for each
target.  Flat fields were obtained by comparing staring images taken at
both high and low airmass.
For photometric calibration of LWS data we compared count
rates to the following known (12.5 and 17.9 $\mu$m) flux densities of
standard stars:  $\alpha$ Lyr, 26.4 Jy and 12.9 Jy; $\sigma$ Lib, 120.7
Jy and 58.9 Jy; $\alpha$ CrB, 3.64 Jy and 1.97 Jy; $\gamma$ Aql, 54.3
Jy and 27.5 Jy. For our MIRLIN data we used the following known (11.7,
12.5, and 20.8 $\mu$m) flux densities and stars: $\gamma$ Aql, 61.7 Jy,
54.3 Jy, and 20.6 Jy; $\beta$ Peg, 313 Jy, 277 Jy, and 107 Jy; $\alpha$
Ari, 62.9 Jy, 55.5 Jy, and 21.1 Jy; $\alpha$ CMi, 59.7 Jy, 52.2 Jy, and
18.8 Jy.  The values are derived from the standard system of
\citet{tok84}. Comparing raw photometry over a range of airmasses let
us find the extinction corrections: 0.12 and 0.35 mag/airmass at 12.5
and 17.9 $\mu$m respectively for the June data; 0.08, 0.10, and 0.40
mag/airmass at 11.7, 12.5, and 20.8 $\mu$m respectively for the
November data.  To maximize the signal-to-noise ratio in the photometry
we used aperture corrections derived from nearby standard star radial 
profiles.

The visible images were obtained while guiding on a nearby star with
sidereal tracking rates in July but non-sidereal rates in
November.  A flat field was obtained by combining images of the blank
twilight sky. Flux calibration and airmass corrections were calculated
by measurements of the \citet{lan92} standard stars SA 107-457, -456
and the PG 1323-086 group in July 2000, and stars SA 98-966, -1002, -L3,
and -L4 in November 2000.

%%%%%%%%%%%%%%%%%%%%%%%%%%
\section{Analysis}

The basic radiometric method to obtain an effective radius $R$ and
geometric albedo $p$ is to solve two equations with these two unknowns,
first done by \citet{all70} and described in detail by \citet{ls89}:

\begin{mathletters}
\begin{eqnarray}
F_{vis}(\lambda_{vis}) & = & 
		{{F_{\odot}(\lambda_{vis})}\over{(r/1{\rm AU})^2}}\ \pi R^2 p\ 
				 {{\Phi_{vis}}\over{4\pi\Delta^2}}, \\
F_{mir}(\lambda_{mir}) & = & 
	\epsilon\!\int\!B_\lambda(T(pq,\Omega),\lambda_{mir}) d\Omega\ R^2\ 
			{{\eta \Phi_{mir}}\over{4\pi\Delta^2}},
\end{eqnarray}
\end{mathletters}

where $F$ is the measured flux density of the object at wavelength
$\lambda$ in the visible (``vis'') or mid-infrared (``mir'');
$F_{\odot}$ is the flux density of the Sun at Earth as a function of
wavelength; $r$ and $\Delta$ are the object's heliocentric and
geocentric distances, respectively; $\Phi$ is the phase function in
each regime; $B_\lambda$ is the Planck function; $\epsilon$ is the
infrared emissivity; $\eta$ is a factor to account for infrared
beaming; and $T$ is the temperature, which is a function of $p$,
surface planetographic position $\Omega$, and the phase integral $q$
which links the geometric and Bond albedos.  For lack of detailed shape
and rotational information -- as is the case for our six objects -- the
modeled body is assumed to be a sphere.

The temperature is calculated using a model of the thermal behavior.
Unfortunately, the thermal inertias are largely unknown so
we use two widely-employed simple models that cover the extremes of
thermal behavior: one for slow-rotators and one for fast.
The former (a.k.a. ``standard thermal model'', STM) applies if the
rotation is so slow (or the thermal inertia so low) that every point on
the  surface is in instantaneous equilibrium with the impinging solar
radiation.  The latter (a.k.a. ``isothermal latitude model'', ILM)
applies if the rotation is so fast (or the thermal inertia so high)
that a surface element does not appreciably cool as it spins away from
local noon and out of sunlight.  The extreme case occurs when
the object's rotation axis is normal to the Sun-object-Earth plane.

There are other parameters to the models: $\epsilon$, $\eta$,
$\Phi_{mir}$, $\Phi_{vis}$, and $q$. Emissivity is close to unity
and we will assume a constant value of 0.9 here.  The beaming parameter
is unity by definition in the ILM. In the
STM, the value is known for only a few asteroids and can range from
approximately 0.7 to 1.2 \citep{har98,har99}. For our applications of
the STM we assume a possible range of $0.75\le\eta\le1.25$, uniformly
distributed. 

For $\Phi$ we assume that the magnitude scales with the phase angle
$\alpha$: $-2.5\log\Phi = \beta\alpha$. In the MIR, the effect is only
loosely constrained but based on earlier work with the STM
\citep{leb86} we assume a range of $0.005$ mag/deg
$\le\beta_{mir}\le0.017$ mag/deg, uniformly distributed. For the ILM
$\beta_{mir}=0$ by definition. In the visible, studied in much more
detail (e.g. by \citet{lb81}), dark asteroids follow
$\beta_{vis}\approx0.035$ mag/deg, and we assume here $0.025$ mag/deg
$\le\beta_{vis}\le0.045$ mag/deg. This coefficient also determines
$q$, but since that has a minor effect on the modeling we
will leave that parameter a constant 0.5, the integral's value for
$\beta_{vis}=0.034$ \citep{all73}.

Our radius and albedo results from this modeling
are shown in Table 3. Since we have as many or more
parameters in the model than data points, useful $\chi^2$-distribution
calculations are impossible. Instead, we found the range of valid $R$
and $p$ by sampling parameter space and declaring a good fit when each
model flux density passed within 1.5-$\sigma$ of its data point. The
``error bars'' in Table 3 actually show the full range (not 1-$\sigma$
limits) of values that yield acceptable fits. For 1999 LE$_{31}$, we
used the average of the modeling results from the two nights of MIR
photometry.

The STM and ILM predict very different color temperatures, providing a
way to differentiate them.  For 2000 DG$_8$, 2000 OG$_{44}$, and 2000
SB$_1$, only the STM provides acceptable fits.
For 2000 PG$_3$, the ILM generally gives poorer
fits and is acceptable only for a
small range of radii and albedos. For 1999 LE$_{31}$ and 2000
HE$_{46}$, we only have one MIR wavelength and thus
cannot find the color temperature.  We note that the ILM albedos are
very small, smaller than for any other Solar System object. This,
combined with the generally poorer fits, implies that the objects are
better interpreted as being slow rotators, although the heliocentric
distance of 1999 LE$_{31}$ may be high enough to make it a borderline case
between the two extremes \citep{sls89}.

Some further caveats are notable. First, the values in Table 3 are
valid in the context of the thermal models; the models represent
extrema of thermal behavior and the ranges in Table 3 do not
describe the systematic errors from the models themselves.  However
these errors are likely to be comparable to the formal errors so
the values are physically meaningful. Second, rotational variation
could corrupt the albedo calculations for the two asteroids with
non-simultaneous MIR and visible photometry, which effectively
increases the albedo errors. Third, our observations of 2000 PG$_3$
took place at $\alpha=59^\circ$, which is beyond the range of any
measured thermal phase function, so we do not know if the standard
formula for $\Phi_{mir}$ mimics reality at such angles.

The radii of our six objects are not unusual in comparison to
cometary nuclei \citep{mhm00} and other, dynamically-similar asteroids
\citep{vee89}, and we now turn our attention to the albedos.  Figure 1
displays the albedos and Tisserand invariants of our six objects as
well as all other, 42 known albedos of NEAs and UAs.
Also plotted are albedos of active cometary nuclei. 

Whereas previously there were only four $T_J<3$ asteroids on the plot,
there are now ten, and all six new ones fall squarely within the
cometary regime of albedos. Moreover, with these new additions we now
see evidence of a trend with $T_J$. Ninety percent of the $T_J<3$
asteroids have comet-like albedos (median albedo is $0.038$),
whereas only two of the 38 $T_J>3$ objects do (median
albedo is $0.215$). An important bias
in the distribution here is that it is easier to discover shinier
asteroids and thus they are overrepresented in the sample, but
nevertheless the difference in distributions is quite stark.

The break in the distributions occurs near the line $T_J=3$, the cutoff
for Jupiter coupling, which suggests a dynamical relationship with this
effect. The low albedos of nearly all $T_J<3$ objects are consistent
with a significant fraction of extinct comets among the
NEA and UA population. Specifically, if 10\% of
NEAs and UAs have $T_J<3$, and 90\% of those have
comet-like albedos, the extinct comet candidate fraction would be
approximately 9\%. This is a lower limit to the actual fraction of 
candidates because of the albedo bias mentioned above.

%%%%%%%%%%%%%%%%%%%%%%%%%%%%%%%%%%%%
\section{Summary}

Using MIR and visible photometry, and employing the
widely-used standard thermal model (STM) for slow rotators, we have derived
new effective radii and geometric albedos for six asteroids in
comet-like orbits; all six have $T_J<3$. We find the following:

$\bullet$ All six objects are dark, as dark as
the albedo spread of cometary nuclei. The radii are also similar
to those of active nuclei. This is consistent with a cometary origin,
as if the asteroids were formerly active comets that lost all
near-surface volatiles.

$\bullet$ For four objects we have photometry at 2 or 3
MIR wavelengths, and the STM yields an excellent description
of the color temperature, better than the fast-rotator model.

$\bullet$ Plotting all 48 known NEA and UA albedos -- including our 6
new ones -- vs. $T_J$ shows a markedly sharp break, virtually a step
function, at the line $T_J=3$.

$\bullet$ Eleven of the 48 objects have comet-like albedos: fully
90\% (9 of 10) of the $T_J<3$ objects, but 
only 5\% (2 of 38) of the $T_J>3$ objects.  The median albedos $\bar p$ and
their r.m.s. scatters are:
\begin{mathletters} \begin{eqnarray}
        \bar p & = & 0.038\pm0.043, {\rm\ for\ } T_J\le3, {\rm\ and} \\
        \bar p & = & 0.215\pm0.147, {\rm\ for\ } T_J>3.
\end{eqnarray} \end{mathletters}

$\bullet$ This disparity in median albedo suggests the
fraction of extinct comets among NEAs and UAs
is significant (at least 9\% are candidates) and
that enough cometary nuclei have sufficiently long physical lifetimes
to survive devolatilization without disintegrating.

It is clear that further surveys of asteroid albedos are necessary.
Specifically, only 4 of the 10 $T_J<3$ objects are NEAs; more members
of that group need to be sampled.  Furthermore, with the recent
explosion in asteroid discoveries and the ready availability of
sensitive MIR detectors, a less biased sampling of albedos should be
undertaken to obtain a confident estimate of the albedo distributions.

%%%%%%%%%%%%%%%%%%%%%%%%%%
\acknowledgments{
We are indebted to Michael Ressler for allowing MIRLIN's use on Keck
and to Varoujan Gorjian for instrument support. The operation of MIRLIN 
is supported by an award from NASA's Office of Space Science.
We thank operators Joel Aycock, Meg Whittle, Wayne Wack, and John
Dvorak for their assistance. We acknowledge the JPL SSD group for their
very useful ``Horizons'' ephemeris program.  We appreciate the help of
Olivier Hainaut in improving this manuscript.  This work was supported
by grants to DCJ from NSF.}

%%%%%%%%%%%%%%%%%%%%%%%%%%

\clearpage

%%%%%%%%%%%%%%%%%%%%%%%%%%

%table 1. orbital elements.

\clearpage
\begin{table}
\begin{center}
\caption{Orbits and Observing Geometry \vskip 5 pt}
\scriptsize
\begin{tabular}{ccccccccc}
\tableline\tableline
Object & $a$ & $e$ & $i$ & $T_J$ & UT Date & $r$ & $\Delta$ & $\alpha$ \\
       & (AU) &  & ($^\circ$) &  & (A.D.2000) & (AU) & (AU) & ($^\circ$)  \\
\tableline
1999 LE$_{31}$ & 8.16 & 0.472 & 152 &  -1.31 & Jun 22 & 5.238 & 5.118 & 11.2 \\
       "       &  "   &   "   &  "  &    "   & Jun 23 & 5.240 & 5.137 & 11.2 \\
       "       &  "   &   "   &  "  &    "   & Jul  2 & 5.267 & 5.311 & 11.0 \\
2000 HE$_{46}$ & 24.3 & 0.903 & 158 &  -1.51 & Jun 23 & 2.526 & 2.470 & 23.5 \\
       "       &  "   &   "   &  "  &    "   & Jul  2 & 2.562 & 2.689 & 22.2 \\
2000 DG$_8$    & 10.8 & 0.793 & 129 &  -0.62 & Nov  8 & 2.319 & 1.728 & 22.9 \\
2000 OG$_{44}$ & 3.88 & 0.581 & 7.33 &  2.74 & Nov  8 & 1.665 & 0.920 & 30.6 \\
2000 PG$_3$    & 2.83 & 0.859 & 20.5 &  2.55 & Nov  8 & 1.065 & 0.929 & 59.1 \\
2000 SB$_1$    & 3.34 & 0.541 & 22.2 &  2.81 & Nov  8 & 1.554 & 0.673 & 25.6 \\
\tableline
\end{tabular}
\tablecomments{$a$ = semimajor axis, $e$ = eccentricity,
$i$ = inclination, $r$ = heliocentric distance at time
of observation, $\Delta$ = geocentric distance at time 
of observation, and $\alpha$ = phase angle at time of
observation.}
\end{center}
\end{table}

%table 2. observing circumstances.

\clearpage
\begin{table}
\tablewidth{0pc}
\begin{center}
\caption{Asteroid Photometry \vskip 5pt}
\scriptsize
\begin{tabular}{ccccccc}
\tableline\tableline
Object & UT Date & UT Time & Wavelength & Flux Density & Meas.  \\
       & (A.D.2000) &      &  ($\mu$m)\tablenotemark{a} & 
	(mJy, Jy, or mag)  & \tablenotemark{b} \\
\tableline
1999 LE$_{31}$ & Jun 22 & 07:11-07:27 & 12.5 & $6.35\pm0.78$ mJy & 2\\
      ''    & Jun 23 & 06:32-06:49 & 12.5 &  $5.65\pm0.71$ mJy & 2 \\
      ''    & Jun 23 & 06:17       & 17.9 & $\le57$ mJy\tablenotemark{c}  & 1\\
      ''    & Jul  2 & 07:32       & 0.65 &  $20.44\pm0.05$ mag & 1   \\
2000 HE$_{46}$ & Jun 23 & 08:01-08:20 & 12.5 &  $29.4\pm1.9$ mJy & 2 \\
      ''       & Jul  2 & 06:16       & 0.65 &  $20.11\pm0.02$ mag & 1 \\
2000 DG$_8$    & Nov  8 & 13:06-13:12 & 11.7 &  $0.382\pm0.013$ Jy & 4 \\
      ''       & Nov  8 & 12:35-13:38 & 12.5 &  $0.411\pm0.037$ Jy & 13 \\
      ''       & Nov  8 & 12:43-13:28 & 20.8 &  $0.71\pm0.15$ Jy & 3 \\
      ''       & Nov  8 & 12:12-12:47 & 0.65 &  $16.826\pm0.016$ mag & 8 \\
2000 OG$_{44}$ & Nov  8 & 06:07-06:12 & 12.5 &  $0.766\pm0.032$ Jy & 3 \\
      ''       & Nov  8 & 06:14-06:20 & 20.8 &  $0.739\pm0.079$ Jy & 4 \\
      ''       & Nov  8 & 06:15-06:17 & 0.65 &  $16.39\pm0.01$ mag & 2\\
2000 PG$_3$    & Nov  8 & 04:39-05:13 & 12.5 &  $0.650\pm0.094$ Jy & 6 \\
      ''       & Nov  8 & 04:43-05:15 & 20.8 &  $0.60\pm0.12$ Jy & 2 \\
      ''       & Nov  8 & 05:11-05:16 & 0.65 &  $17.857\pm0.013$ mag & 2 \\
2000 SB$_1$    & Nov  8 & 11:07-11:09 & 11.7 &  $1.112\pm0.035$ Jy & 2 \\
      ''       & Nov  8 & 11:02-11:48 & 12.5 &  $1.139\pm0.061$ Jy & 8 \\
      ''       & Nov  8 & 11:11-11:40 & 20.8 &  $1.48\pm0.35$ Jy & 3 \\
      ''       & Nov  8 & 11:15-11:18 & 0.65 &  $16.27\pm0.01$ mag & 2 \\
\tableline
\end{tabular}
\tablenotetext{a}{``$0.65$'' refers to R band.}
\tablenotetext{b}{``Meas.'' signifies the number of measurements that
were used to calculate the value in the ``Flux Density'' column.}
\tablenotetext{c}{This is a 3-$\sigma$ upper limit.}
\end{center}
\end{table}

% table 3. radii. and albedos.

\clearpage
\begin{deluxetable}{ccccc}
\tablecolumns{6}
\tablewidth{0 pc}
\tablecaption{Effective Radii and Geometric Albedos\tablenotemark{a}}
\tablehead{
\colhead{Object} & \multicolumn{2}{c}{Slow Rotator Model} & \multicolumn{2}{c}{Fast Rotator Model} \\
\cline{2-3}   \cline{4-5} 
\colhead{}       & \colhead{Eff. Radius (km)} & \colhead{Geom. Albedo} & \colhead{Eff. Radius (km)} & \colhead{Geom. Albedo}} 
\startdata
1999 LE$_{31}$     & $9.05^{+4.04}_{-2.71}$ & $0.031^{+0.030}_{-0.020}$ & $23.37^{+2.93}_{-2.67}$ & $0.0041^{+0.0018}_{-0.0014}$ \\
2000 HE$_{46}$     & $3.55^{+1.10}_{-0.78}$ & $0.023^{+0.021}_{-0.013}$ & $6.36^{+0.29}_{-0.28}$  & $0.0067^{+0.0022}_{-0.0021}$ \\
2000 DG$_8$        & $8.64^{+2.26}_{-1.83}$ & $0.027^{+0.022}_{-0.015}$ & \tablenotemark{b}       & \tablenotemark{b}            \\
2000 OG$_{44}$     & $3.87^{+0.50}_{-0.40}$ & $0.038^{+0.018}_{-0.017}$ & \tablenotemark{b}       & \tablenotemark{b}            \\
2000 PG$_3$        & $3.08^{+1.42}_{-0.95}$ & $0.021^{+0.031}_{-0.017}$ & $3.49^{+0.21}_{-0.19}$\tablenotemark{c} & $0.015^{+0.007}_{-0.009}$\tablenotemark{c}   \\
2000 SB$_1$        & $3.57^{+0.92}_{-0.62}$ & $0.019^{+0.015}_{-0.010}$ & \tablenotemark{b}       & \tablenotemark{b}            \\
\enddata
\tablenotetext{a}{With full ranges of acceptable values 
(not 1-$\sigma$ limits).}
\tablenotetext{b}{The fast model gives unacceptable fits for these objects.}
\tablenotetext{c}{The fast model gives generally poorer fits and is
acceptable for only a small spread of radii and albedos.}
\end{deluxetable}

\clearpage

\figcaption[fig1.eps]{Plot of Tisserand invariant vs. all known
geometric albedos for $T_J<3$ NEAs and UAs (green circles),
$T_J>3$ NEAs and UAs (blue hourglasses), and comets 
(red squares). The six objects presented in this Letter are marked as filled
circles with estimated 1-$\sigma$ errors. A heavy vertical 
line marks the dynamical boundary
$T_J=3$ (see \S 1).  Of the 48 total asteroids plotted, 11 have
comet-like albedos, or 23\%, but fully 90\% (9 of 10) of those
with $T_J<3$ have comet-like albedos. This is consistent
with a cometary origin for those asteroids and a significant cometary
contribution to that dynamical group of asteroids. The plotted data
were obtained from this work,
\citet{mor76}, \citet{cj77},  \citet{leb78}, \citet{leb79},
\citet{leb81}, \citet{gre85}, \citet{tg87},  \citet{bel88},
\citet{vee89}, \citet{ted92}, \citet{ho95},
\citet{mot97}, \citet{pra97}, \citet{har98}, \citet{tho00},
\citet{har99}, \citet{fer99}, \citet{del00}, and \citet{fer01}.}

\end{document}